\documentclass[english,a4paper,citeautoscript,prb,reprint,amsmath,amssymb,superscriptaddress,nofootinbib]{revtex4-1}
\usepackage{preamble}
\begin{document}

\title{Liquid-liquid phase separation, biomolecular condensates, puncta,  non-stoichiometric supramolecular assemblies, membraneless organelles, and bacterial chemotaxis are best understood as emergent phenomena with switch-like behaviour}

\author{Richard P Sear}
\affiliation{Department of Physics, University of Surrey, Guildford, Surrey GU2 7XH, United Kingdom}
\email{r.sear@surrey.ac.uk}

\begin{abstract}
%\large
Liquid-liquid phase separation (LLPS) is currently of great interest in cell biology. LLPS is an example of what is called an emergent phenomenon -- an idea that comes from condensed-matter physics. Emergent phenomena have the characteristic feature of having a switch-like response. I show that the Hill equation of biochemistry can be used as a simple model of strongly cooperative, switch-like, behaviour. One result is that a switch-like response requires relatively few molecules, even ten gives a strongly switch-like response. Thus if a biological function enabled by LLPS relies on LLPS to provide a switch-like response to a stimulus, then condensates large enough to be visible in optical microscopy are not needed.
\end{abstract}

\maketitle

\section{Introduction}

\begin{figure}[htb!]
  \begin{center}
     \includegraphics[width=7.5cm,angle=270]{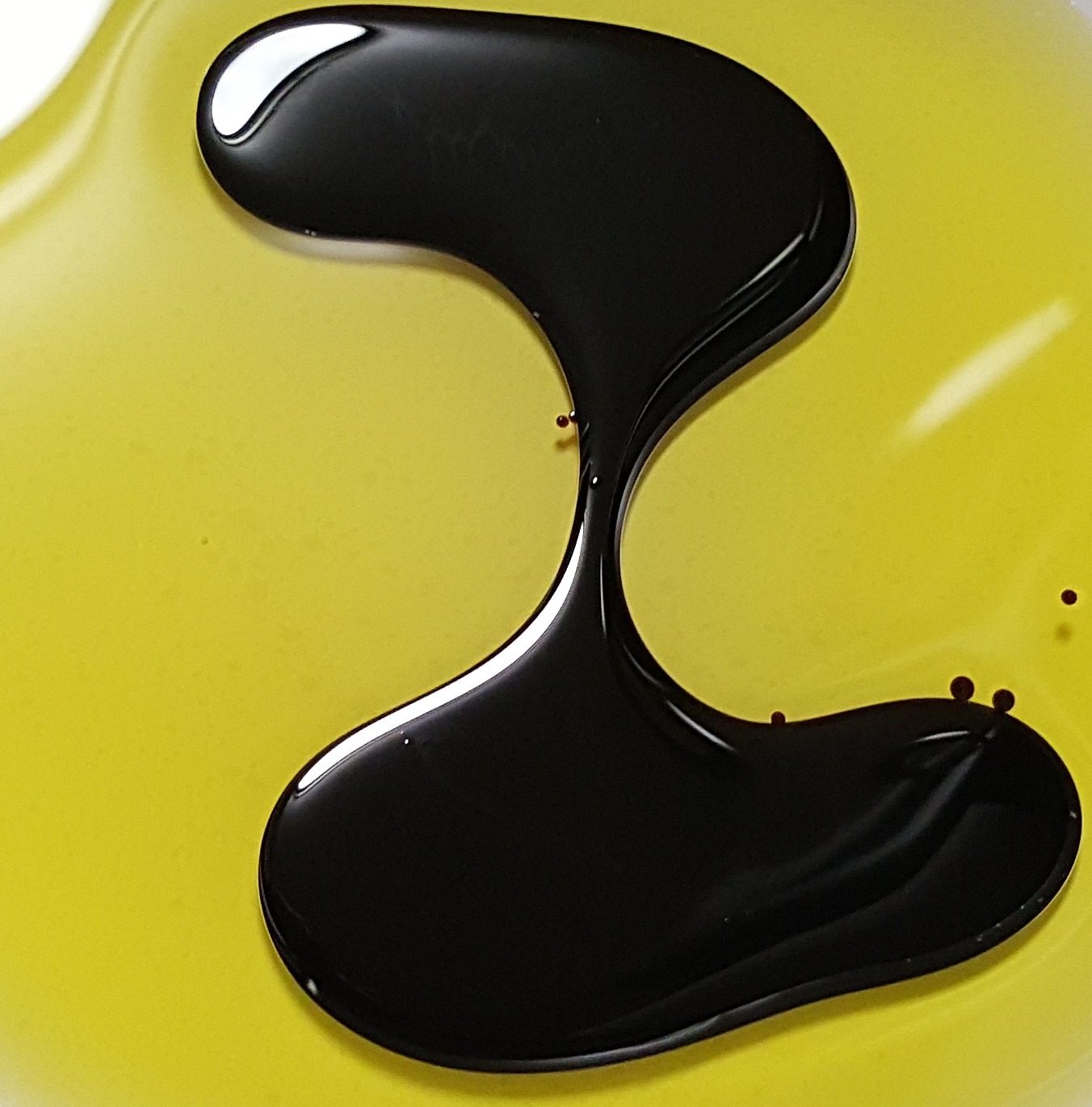}
      \caption{Liquid/liquid phase separation of balsamic vinegar (black) and olive oil (yellow). From \href{https://commons.wikimedia.org/wiki/File:Balsamoil_Protean.jpg}{Wikimedia, author Drazmoyde}, made available under a
       \href{https://creativecommons.org/licenses/by/4.0/deed.en}{Creative Commons Attribution 4.0 International license}.
}
      \label{fig:LLPS}
  \end{center}
\end{figure}

In cell biology, there has been an explosion of interest in what is variously called \lq\lq liquid/liquid phase separation\rq\rq~(LLPS) \cite{wu20}, \lq\lq biomolecular condensates\rq\rq~\cite{banani17}, \lq\lq puncta\rq\rq ~\cite{bienz14,sear07},  \lq\lq non-stoichiometric supramolecular assemblies\rq\rq~\cite{pancsa19} or \lq\lq membraneless organelles\rq\rq~\cite{meszaros19}.
Of the five terms, four have been recently invented by biologists, while LLPS is a well-established term in thermodynamics \cite{hansen13, degennes13}. LLPS is illustrated in Figure \ref{fig:LLPS}, where liquid olive oil (yellow), and liquid vinegar (black), have phase separated.
There are a number of good recent reviews, see for example \cite{banani17,meszaros19,swain20,chen20,ning19,wu20,you19,bienz14,bienz20,pancsa19}.

%Here, I want to do two things: Firstly, I want to carefully define what physical scientists and engineers mean by LLPS. There is large literature  on understanding LLPS, but it is written by physical scientists and engineers, for physical scientists and engineers. It therefore makes hard reading for many biologists. So I want to discuss some of what is known about LLPS, focusing on what I believe are the aspects most relevant to biological function, and in a way that I hope is accessible to biologists. I hope this will help biologists apply this understanding to better understand LLPS in cells.

Here, I want to point out to biologists that LLPS is just one example of a broad class of phenomena, called emergent phenomena \cite{laughlin14,anderson72,hansen13,pancsa19}. I carefully define emergent phenomena below, but in brief, they are phenomena that can occur (emerge) {\em only} when many molecules (or other objects) interact. Other examples of emergent behaviour in cell biology are known \cite{duke99,sourjik04}, as are other examples in biology. For example the collective behaviour of flocks of birds and swarms of insects are emergent phenomena \cite{bialek12}. The idea of emergent phenomena is arguably the most interesting and productive idea to arise in physics in the second half of the twentieth century. About one third of the Nobel Prizes in Physics awarded from 1970 to 2000, were for aspects of emergent phenomena.
Now in the early twenty-first century, we find that there are many examples of what looks like an emergent phenomenon in cell biology. It is the perfect time to apply what we know about emergent phenomena in non-living systems \cite{laughlin14,anderson72,hansen13,pancsa19}, to cell biology.

\section{Defining liquid-liquid phase separation (LLPS)}

I start by defining what a liquid is.
The definition of a liquid is simple. It is a phenomenological not a molecular definition. Liquids are distinguished from solids by the fact that when a force is applied to a liquid it flows, whereas solids do not. Thus parts of the cytoplasm or nucleoplasm that exhibit flow, are liquids.

%The definition has two parts. The first is that a liquid must be able to flow, you must be able to pour a liquid. This is sometimes expressed by saying that the liquid takes the shape of its container. This distinguishes liquids from solids. The second part is that liquids are dense and so have a definite volume. The contents of cells are hard to compress, and flow (slowly) so they are liquid.

Now that we know what a liquid is, then LLPS is just two distinct coexisting liquids. An example is shown in  Figure \ref{fig:LLPS}. Both the balsamic vinegar and the olive oil satisfy the definition of a liquid, and they coexist.
The influential work of Brangwynne and coworkers found that P granules in the cytoplasm of {\em Caenorhabditis elegans} behaved much like droplets of balsamic vinegar in olive oil, the granules flowed and coalesced \cite{brangwynne09}.

As the definition of liquid is phenomenological not based on molecular behaviour, it goes against how a liquid is defined, to use molecular behaviour to decide if something is a liquid or not. So an experimental technique such as Fluorescence Recovery After Photobleaching (FRAP) that studies dynamics of molecules, cannot determine if a region in a cell is liquid or not.
If a region of a cell with a high concentration of a particular protein is characterised via a technique such as FRAP that studies molecular properties, it is more sensible to refer to it as a biomolecular condensate or non-stoichiometric assembly etc, rather than LLPS.

%In the remainder of this article, I want to highlight some key features of LLPS from a physicist's point of view, and point out where I think they may have relevance to biological function. Not everything observed in a cell is required for function so I do not want to just assume that LLPS is required for function, I would like to look at the features of LLPS that look useful.

\section{LLPS is an an example of an emergent phenomenon}

LLPS is an example of an emergent phenomenon \cite{laughlin14,anderson72,hansen13,pancsa19}.
By definition, emergent phenomena are phenomena that {\em only} appear when many molecules interact and act as one. One or two molecules cannot show emergent phenomena.
Emergent behaviour is also sometimes referred to as behaviour where \lq\lq more is different\rq\rq, following an iconic (amongst physicists) article \cite{anderson72} by arguably the most distinguished condensed matter physicist of the postwar period, Philip Anderson.

In the case of liquids and LLPS, the most obvious emergent (particularly if microscopy is the technique) phenomenon is the ability to flow. A single molecule, or a dimer, cannot flow but a condensate of hundreds or more molecules can. This flow can then be described by an equation, the Stokes equation, which applies to liquids but not to single molecules.

Within the Stokes equation, the liquid is described solely by its viscosity, which describes how fast or how slowly a liquid flows when pushed. A characteristic and fundamental feature of emergent properties (here viscosity), is that although they are determined by the molecular properties, they give very little information on these molecular properties \cite{laughlin14}. For example, the viscosity of droplets of the
{\em Caenorhabditis elegans} protein LAF-1, has
been estimated to be $34\pm 5$~Pa~s \cite{elbaum15}. This viscosity is determined by the molecule LAF-1, however, the number of different molecular interactions that can give rise to a viscosity of 34~Pa~s is infinite, and so measuring the viscosity tells you almost nothing about LAF-1.

The ability to flow is just one of the emergent phenomena of LLPS, there are others. For example, when the new liquid forms, a new physicochemical environment emerges inside the liquid, which will have new affinities for biomolecules, and so an ability to sequester biomolecules emerges.

%\subsection*{\large LLPS is emergent behaviour resulting in flow}

%The distinguishing feature of LLPS, as opposed to say for example ferromagnetism which is also a emergent phenomenon, is the ability to flow of the liquids. However, although Setru and coworkers look at functional behaviour in LLPS that relies on flow \cite{setru20}, this is a rare example. In most cases there is no evidence that the ability to flow is required for biological function.

\subsection{A biological function as an emergent phenomenon}

If a biological function of a set of biomolecules in a cell
relies on an emergent phenomenon then the function must obey the standard rules for emergent phenomena \cite{laughlin14,anderson72}. These are that the function relies on many interacting biomolecules, it cannot be done by one or a few, and that it will be independent of many details of these molecules. This independence of many details of the molecular interactions has advantages and disadvantages. It has the advantage that we do not need to know these details to understand how the function works, which can greatly simplify what otherwise may be some complex cell biology. The disadvantage is that studying the functions tells you little about the underlying biomolecules.

%\begin{figure*}[htb]
%  \begin{center}
%     \includegraphics[width=14.8cm]{chemotaxisEcoli.png}
%      \caption{\large Schematic illustrating difference between, from left to right, monomeric receptors, cooperative dimeric recpetors, and a cluster showing emergence. 
%      showing membrane plus transmembrane receptors.
%      Our model of {\em E.~coli} sensing of aspartate concentrations assumes that the transmembrane receptors have two states: active (purple) and inactive (yellow). Binding of a molecule, eg aspartate (shown as diamond) favours the inactive state. The emergence comes from the fact that the receptors are in a two-dimensional array of thousands of receptors \cite{sourjik04b}, and that receptors interact with neighbouring receptors. The quantitative data \cite{sourjik04} on sensing require cooperative interactions between neighbouring receptors such that when one receptor flips to the inactive conformation, this makes neighbouring receptors more likely to flip to the inactive conformation. Created with Biorender.
%}
%      \label{fig:chemotaxis}
%  \end{center}
%\end{figure*}

\section{LLPS is not the only example of an emergent phenomenon in cells}

The role of emergent phenomena in {\em E.~coli} chemotaxis has been extensively studied \cite{duke99,sourjik04}. I think those studying LLPS can learn from this work. So I will briefly review it, highlighting the features in common with LLPS.

\subsection{{\it E.~coli} chemotaxis}

Bacteria such as {\em E.~coli} can sense gradients in the concentration of food molecules such as aspartate, with impressive accuracy \cite{sourjik04,bray04}. This enables them to swim to regions of higher food concentration, which is called chemotaxis. The role of the emergent phenomenon here is to provide the cooperativity needed to increase the sensitivity with which the bacteria sense their environment.

The receptors in {\em E.~coli} chemotaxis form long-lived assemblies in which they interact cooperatively \cite{sourjik04,bray04}. The receptors do not assemble cooperatively, or flow. However, the emergent behaviour that drives signal transduction is known to be analogous to that in LLPS.
Duke and Bray modelled {\em E.~coli} chemotaxis \cite{duke99} using a condensed-matter physics model (called the Ising model) also used to model LLPS \cite{chandler_book}. 
Thus, if the biological function of LLPS is to provide cooperativity that it is very closely analogous to the behaviour seen in {\em E.~coli} chemotaxis, despite the absence of liquid behaviour in the chemotaxis receptors.

%So the {\em E.~coli} chemotaxis receptors and some of the cell signalling systems showing LLPS may be examples of systems that may look very different to microscopy but where analogous emergent behaviour is used to achieve function. 

\section{Ten molecules is enough for strongly emergent behaviour}

Emergent behaviour is an extreme example of cooperative behaviour. This is true in the sense that emergence is defined to mean that there is no limit to the strength of the cooperativity, because there is no limit to the number of molecules that can interact and cooperate.
However, in practice even relatively small numbers of molecules, tens or fewer, can often result in strong cooperativity \cite{imry80,binder03}. So even if a biological function relies on a set of biomolecules capable of emergent behaviour, it may be that in practice the function is performed by tens of molecules, not thousands.
I will try and show this, by following Sourjik and Berg, and using the Hill equation \cite{sourjik04}. 
%They applied the Hill equation to {\em E.~coli} chemotaxis \cite{sourjik04}).

\subsection{Understanding emergent biological functions using the Hill equation of biochemistry}

The Hill equation was originally developed to study the cooperative oxygen-binding behaviour of the four subunits of haemoglobin \cite{hill10,weiss97}. Note that as only a maxium of four subunits can cooperate, this is not an emergent phenomenon.

\begin{figure}[htb]
  \begin{center}
     a)
     \includegraphics[width=7.5cm]{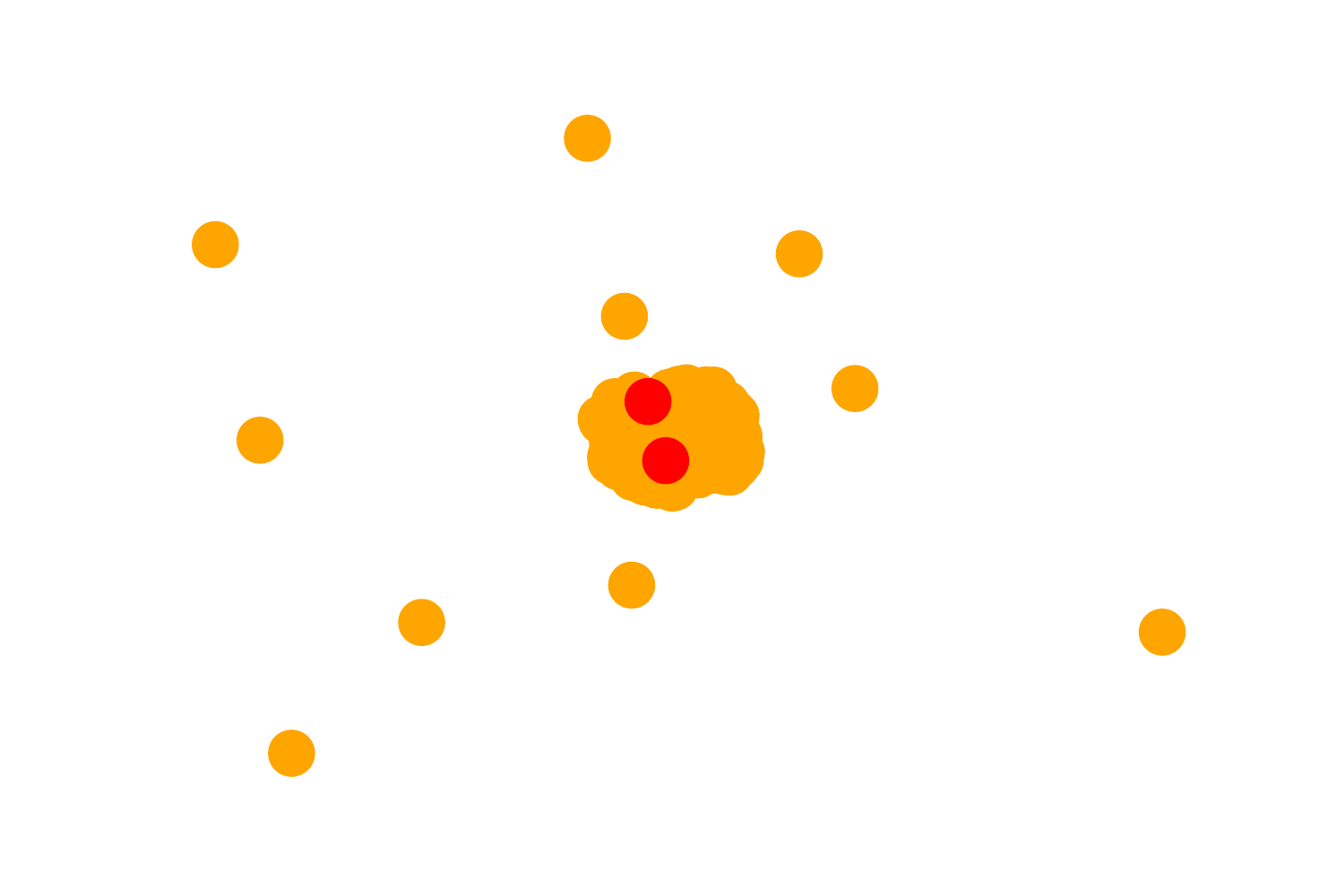}
     b)\includegraphics[width=7.5cm]{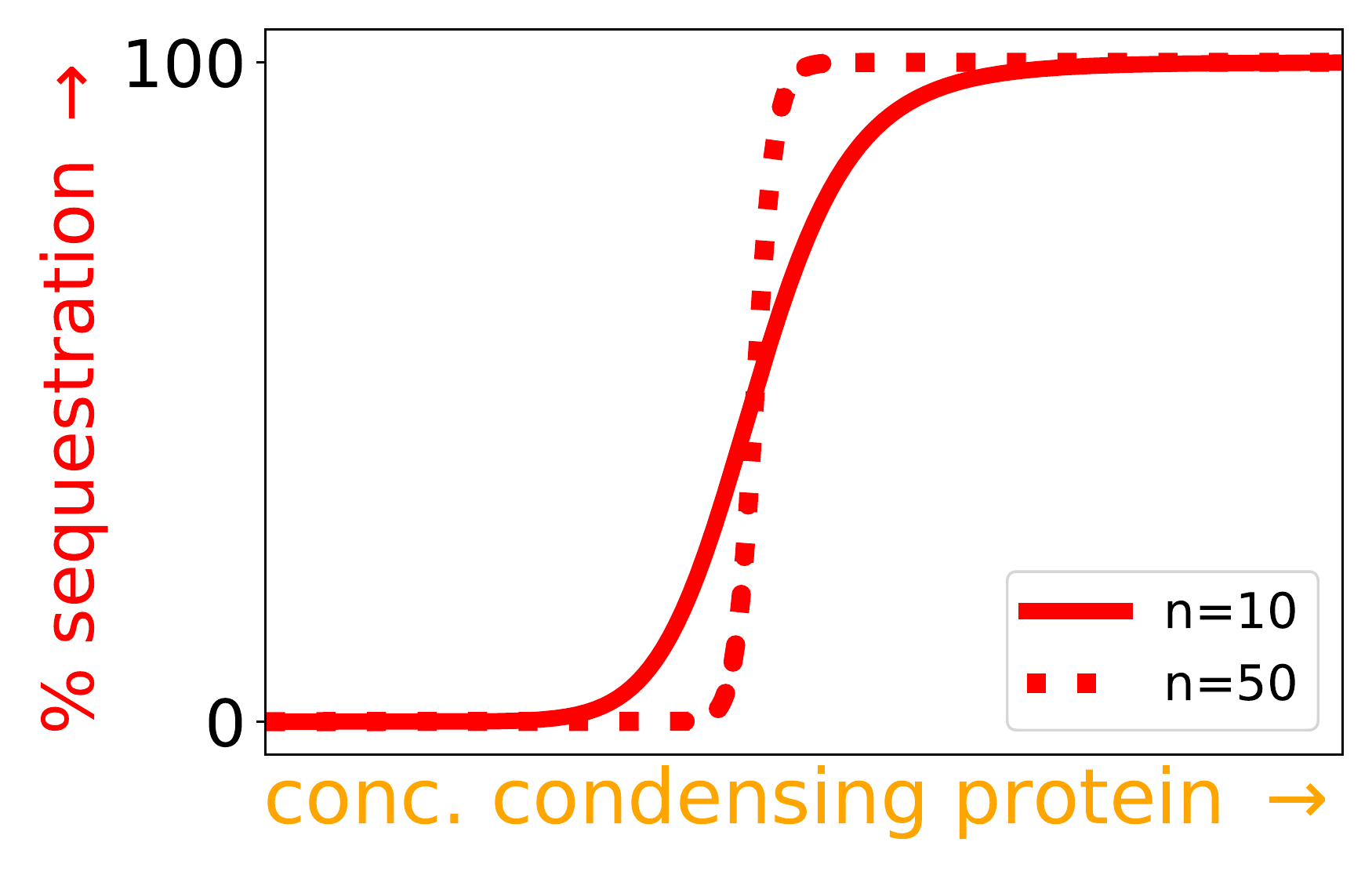}
      \caption{\
      a) A schematic showing a condensate (of orange molecules) which has recruited a target species (red). If recruitment requires multivalent bonding of the red species by more than one neighbouring orange species, then recruitment will be an emergent phenomenon. The multivalent binding then only emerges as the condensate concentrates the orange species.  b) Plots of the fraction of a target sequestered as a function of the concentration of the condensing protein. Both curves are from the Hill equation but with Hill coefficients $n=10$ (solid) and $n=50$ (dashed).
   %  b) Plot of data and fit from Hill's original 2010 \cite{hill10} analysis of O$_2$ binding to heamoglobin. Blue triangles are data from Barcroft and Camis (data is set number III in \cite{hill10}), curve is Hill's fit with Hill coefficient $n=2.49$.
}
      \label{fig:ising}
  \end{center}
\end{figure}

It is perhaps easiest to consider a specific example, so let us look at recruitment of a target biomolecule (shown in red in Figure \ref{fig:ising}) by another species of biomolecule (shown in orange), that condenses. I assume that recruitment relies on formation of a condensate, because it relies on multivalent interactions between the target and the condensing molecules. Interactions that are only present when the condensing molecules are locally at a high concentration.

The Hill equation applies to a very simple (and so highly approximate) model chemical equilibrium. For recruitment, the chemical equilibrium is between free condensing proteins, $C$, free target protein, $T$, and condensates containing $n$ $C$ proteins and target protein, $TC_n$,
$$
T + n C \rightleftharpoons TC_n
$$
%If $C$ is replaced by haemoglobin, $T$ replaced by oxygen, and
For $n=4$, this is
a standard model for oxygen binding to haemoglobin,
but if we allow $n$ to take any value, it can model emergence.

If the function we are interested in is sequestration of the target $T$, then the figure of merit for the function is
$$
%\theta=
\begin{array}{cc}
\mbox{fraction of target $T$}  \\
\mbox{in condensates}
\end{array}=
\frac{1}{1+K_d/[C]^n}
$$
which is the Hill equation \cite{hill10,weiss97};
$K_d$ is a dissociation constant.
%Note that it applies to the chemical equilibrium above, in which case $n$ is the number of molecules.
%However, Hill used $n$ as a fitting parameter, then it is typically called the Hill coefficient, and the best-fit value is typically not precisely equal to the number of molecules, for example, it is less than four for haemoglobin, see Figure \ref{fig:ising}b).
The concentration of sequestering molecules needed to sequester the target only depends on the value of the dissociation constant $K_d$ --- it is a characteristic of emergent phenomena that molecular details only matter in so far as they contribute to the value of one or a few parameters.

The fraction of a target recruited by a condensate of ten molecules is plotted in Figure \ref{fig:ising}. 
%Note the characteristic sigmoidal shape of the fraction of the target recruited, as the concentration of the condensing protein is varied. Also
Note that the onset of sequestration is relatively sharp even for $n=10$ molecules, and is very switch-like
for $n=50$. It is characteristic feature of emergent phenomena that behaviour can be switch-like \cite{laughlin14,anderson72,hansen13,pancsa19}.

In the language of the Hill equation, the formation of even a small condensate has a very large value of the Hill coefficient $n$, and that the Hill coefficient increases with size. As LLPS is a true emergent phenomenon, the Hill coefficient can become arbitrarily large.

%It is true, but much rarer, that what looks like conventional equilibrium LLPS, can also form via very different mechanisms
%But far from thermodynamic equilibrium, coexisting liquid-like objects can also form via much more exotic mechanisms \cite{cates15}.

%So even far from equilibrium, where the molecular behaviour may be quite exotic, it still, I think, makes sense to keep the definition of a liquid phenomenological not molecular. And this holds for LLPS too, if both liquids flow, and if like domains coalesce, then we have LLPS.

%The third complication is that both the cytoplasm and the nucleoplasm contains thousands of different biomolecule species. This makes the coexisting liquids very complex, but we do have some ability to model and understand this complexity \cite{sear03,riback20,??}.

\section{Conclusion}

There has been an explosion of interest in LLPS/\lq\lq biomolecular condensates\rq\rq/\lq\lq non-stoichiometric supramolecular assemblies\rq\rq, $\ldots$ \cite{banani17,meszaros19,chen20,ning19,wu20,you19,bienz14,bienz20,pancsa19}. The ability of these condensates/assemblies to grow very large (eg when a protein is overexpressed) suggests that they are an example of an emergent phenomenon. This is an exciting possibility, emergent behaviour has already been shown to be a powerful way of understanding {\em E.~coli} chemotaxis \cite{duke99}, and it may be equally useful in understanding the many systems now known to have biomolecular condensates.

Focusing on the cooperative nature of emergent behaviour will, I think, be more useful than focusing on the liquid-like nature of the condensates. Other than in work of \cite{setru20}, the liquid-like nature of the condensates has not been shown to be directly relevant to function.
If cooperativity is directly functional, it is worth noting that, as we showed above using the Hill equation, strong cooperativity is achieved with only ten or so molecules.
Thus, condensates/assemblies of ten molecules may be perfectly functional, although hard to detect using conventional microscopy.

\section{Acknowledgements}

It is a pleasure to acknowledge M. Bienz, X. Darzacq and D. Frenkel for helpful discussions. Figure 1 is from
Wikimedia, author Drazmoyde.

%\section{Competing interests}
%No competing interests declared
%\medskip
%\bibliographystyle{ieeetr}
\bibliography{liquids_elife}

\end{document}